\documentclass[aps,prb,showpacs,twocolumn,superscriptaddress,amsmath,amssymb,floatfix]{revtex4}
\usepackage{tabularx}
\usepackage{bm}
\usepackage{epsfig,subfigure}
\usepackage{multirow}
\usepackage{graphicx}
\usepackage{color}
\usepackage{amsfonts}
\usepackage{exscale}
\usepackage{amsbsy}
 \usepackage[colorlinks,urlcolor=blue,linkcolor=blue,anchorcolor=blue,citecolor=blue]{hyperref}

\begin{document}
\title{Obtaining topological degenerate ground states  by the density matrix renormalization group}

\author{Yin-Chen He}
\affiliation{Department of Physics, State Key Laboratory of Surface Physics and Laboratory of Advanced Materials, Fudan University, Shanghai 200433, China}

\author{D. N. Sheng}
\affiliation{Department of Physics and Astronomy, California State University, Northridge, California 91330, USA}

\author{Yan Chen}
\affiliation{Department of Physics, State Key Laboratory of Surface Physics and Laboratory of Advanced Materials, Fudan University, Shanghai 200433, China}

\date{\today}


\begin{abstract}
We develop the density matrix renormalization group approach to
 systematically identify  the topological order of the quantum spin liquid (QSL) through adiabatically obtaining different topological degenerate sectors of the QSL on an infinite cylinder. 
As an application,  we study  the anisotropic  kagome Heisenberg model known for hosting a Z$_2$ QSL, however no numerical simulations have been able
to access all four sectors before. 
We obtain the complete set of four  topological degenerate ground states distinguished by the presence or absence of
the spinon and vison quasiparticle line, which fully characterizes  
the topological nature of the quantum phase.  We have also studied the kagome Heisenberg model, which has recently attracted a lot of attention.
We find  two topological sectors accurately and also estimate various properties of the other topological sectors, where 
the larger correlation length is found indicating the possible proximity to another phase.
\end{abstract}

\maketitle
\section{Introduction}
The quantum spin liquid (QSL), a state which does not break any lattice  or spin-rotational symmetry 
at zero temperature, has attracted much attention in the
past twenty years. \cite{Balents2010}  Different from a trivial disordered state, the QSL possesses topological order~\cite{Wen1990, Wen2004} with the  deconfined and
fractionalized quasiparticles obeying the  anyonic braiding statistics. The physics of the QSL may also have an implication
for understanding the high temperature superconductivity. \cite{Anderson1987} Since Anderson first proposed the resonating valence bond (RVB) state for the 
triangular Heisenberg magnet, \cite{Anderson1973} the debate on whether a QSL is
a realistic quantum state in two dimensional (2D) systems has never ceased. In recent years, there is growing experimental \cite{Kagawa2005, Kurosaki2005, Helton2007, Lee2007, Mendels2007, Okamoto2007,Yamashita2008, Olariu2008, Wulferding2010, Imai2011, Jeong2011, Clark2013, Han2012} and theoretical \cite{Misguich1999,Moessner2001, Balents2002, Sheng2005a, Isakov2006, Isakov2011, Jiang2008, Meng2010, Yan2011, Wang2011, Jiang2012a,  Jiang2012b, Depenbrock2012, Nishimoto2013} evidence  supporting the existence of the QSL in realistic materials or contrived model systems.

Theoretically, frustrated magnetic interactions may lead to a QSL. Those systems impose serious difficulties for theoretical studies, where  analytical methods are under development and the quantum Monte Carlo method  is usually  not applicable due to the sign problem. To tackle these problems, 
the slave particle formalism \cite{Read1991, Wen1991, Lee2005, Wang2006, Ran2007, Iqbal2011,Clark2011, Ruegg2012, Senthil2000} has been 
developed and 
different  model Hamiltonians have been constructed, \cite{Moessner2001, Rokhsar1988, Misguich2002a, Balents2002, Motrunich2002, Levin2003, Kitaev2003, Hermele2004,Ruegg2012}
which give many insights for the properties of the QSL.
However, these studies still can not provide concrete predictions regarding the existence of the QSL in more realistic quantum systems. 
The  development of
the density matrix renormalization group (DMRG) \cite{DMRG} and the tensor network approach \cite{PEPS, Wang2011, MERA, Evenbly2012, Wang2013} has opened a new route to study the QSL in general magnetic systems; in particular,  the
accurate DMRG studies have provided extensive evidence for a possible gapped Z$_2$ QSL for the kagome lattice Heisenberg  model. \cite{Yan2011, Jiang2012b, Depenbrock2012}
However, the  variational Monte Carlo \cite{Ran2007, Iqbal2011} 
study of the same system suggests a possible gapless QSL, which appears to be
more consistent with experimental observations.  \cite{ Helton2007, Wulferding2010, Imai2011, Jeong2011, Clark2013}
Recently, it has been suggested \cite{Wang2013} that both a small correlation length of one topological sector \cite{Yan2011, Depenbrock2012} and the
positively quantized  topological entanglement entropy  used in identifying the $Z_2$ QSL \cite{Jiang2012b, ssg2013h} may not be sufficient 
to fully establish the nature of the quantum state limited by  the range of system sizes being studied.
Therefore, it is crucial to find other topological sectors, which may lead to a full understanding of the topological nature 
of  QSL through extracting  the modular matrix. \cite{Wen1990, Zhang2012, Cincio2013, Zaletel2013, Zhu2013}

DMRG has been proven powerful in solving the ground state of quasi-one dimensional frustrated magnets,  \cite{Jiang2008, Meng2010, Yan2011, Jiang2012a,  Jiang2012b, Depenbrock2012, Nishimoto2013} however it can not directly obtain excited state accurately for larger systems, especially on a torus geometry where the topological degeneracies exist. In a recent work, \cite{Cincio2013} Cincio and Vidal showed that the topological degeneracy can be studied in an infinite cylinder, and they can obtain topological degenerate ground states by using random initial conditions in the infinite DMRG simulation. \cite{McCulloch2008}
In this paper, we propose a systematical  and controlled approach based on  the DMRG calculations
to find different topological degenerate sectors of the quantum system on an infinite cylinder. In Sec. \ref{sec:top_deg}, we briefly review the origin of topological degeneracy and derive topological degeneracies of a Z$_2$ as well as double-semion QSL phases. We show that different topological degenerate ground states differ from each other by certain type of quasiparticle (spinon or vison for the QSL) lines threaded in the system, and these states can be tuned into each other by inserting flux. In Sec. \ref{sec:algorithm}, we outline the general numerical scheme
of the algorithm, which are based on the origin of topological degeneracy. We show that, to obtain topological degenerate ground states in QSL, two operations can be implemented in the DMRG simulation: (1) creating edge spinon. (2) adiabatically inserting  $2\pi$ flux. In Sec. \ref{sec:winding_number}, we discuss the relation between winding number and spinon line. We also propose a simple method to determine the presence of spinon line by observing the entanglement spectrum. In Sec. \ref{sec:easy_axis}, we apply the method to the anisotropic easy axis kagome Heisenberg model (EAKM), which is shown to host a Z$_2$ QSL theoretically \cite{Balents2002} and numerically. \cite{Sheng2005a,Isakov2006,Isakov2011}
We successfully find four topological degenerate states and calculate various quantities to show that they are four distinct states. Further, in Sec. \ref{sec:KHM}, we apply our method to nearest neighbor kagome Heisenberg model (KHM). For the
KHM, we have  only identified two topological sectors, however, our results may 
provide  a good approximation on the properties of other topological sectors.
We find that the states in the new topological sectors  have a much larger correlation length, which demands future study.

\section{Topological degeneracy: Z$_2$ and double-semion quantum spin liquid\label{sec:top_deg}}
\begin{figure}[ht]
\centering
\includegraphics[width=0.35\textwidth]{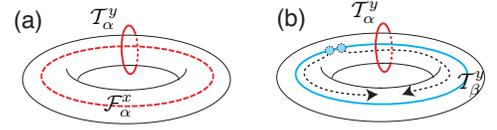} \caption{(color online) \label{fig:top_deg}(a)  Topological sectors created by threading flux. (b) Topological sectors created by dragging out an anyon line.}
\end{figure}

The topological degeneracy originates from the existence of the fractionalized quasiparticles and their
 anyonic braiding statistics. Following Ref. \onlinecite{Oshikawa2006}, we will give an exact derivation of the topological degeneracy in Z$_2$ and double-semion QSL. \cite{Levin2003} Either Z$_2$ or double-semion QSL supports two kinds of fractionalized excitations, spinon and vison. To deduce the topological degeneracies, we should first define the Wilson loop operator $\mathcal{T}_{s}^{x(y)}$ ($\mathcal{T}_{v}^{x(y)})$ (Fig. \ref{fig:top_deg}), which creates a pair of spinons (visons), then winds them along $x$ ($y$) direction, and finally annihilates them. If the system has a gap and we drag the quasiparticles slowly enough, the system will get back into the ground state after the annihilation of quasiparticles. Therefore, no matter whether there is ground state degeneracy, we can always find a ground state $|\psi_0\rangle$ which is the eigenstate of the $\mathcal{T}_{s}^y$ and $\mathcal T_v^y$, satisfying $\mathcal{T}_{s}^y|\psi_0\rangle=\alpha_s|\psi_0\rangle$ and $\mathcal{T}_{v}^y|\psi_0\rangle=\alpha_v|\psi_0\rangle$. Here $\alpha_{s(v)}$ are  non-universal numbers, for simplicity we just take them to be $1$. In the following, we will use two ways to derive that the system also has three other topological degenerate ground states $|\psi_s\rangle$, $|\psi_s\rangle$, $|\psi_{sv}\rangle$, whose corresponding eigenvalues are $(-1,1)$, $(1,-1)$ and $(-1,-1)$:
 \begin{align}
\psi_0&= |\mathcal{T}_s^y= 1, \mathcal T_v^y= 1\rangle, \quad \psi_s= |\mathcal{T}_s^y= -1, \mathcal T_v^y= 1\rangle  \nonumber \\  \psi_v&= |\mathcal{T}_s^y= 1, \mathcal T_v^y= -1\rangle, \,\psi_{sv}= |\mathcal{T}_{s}^y= -1, \mathcal T_v^y= -1\rangle.
\end{align}
 These topological degenerate ground states, defined by eigenstates of the Wilson loop operator along $y$ direction, are the minimal entangled states introduced in Ref. \onlinecite{Zhang2012}. 

One way to derive the topological degeneracy is using the existence of the deconfined fractionalized quasiparticles. Similar 
to the fractional quantum Hall effect, one can insert flux $\theta$ in the torus along $x$ direction. We call the
corresponding operators  $\mathcal F_{s(v)}^x(\theta)$. Then, the spinon (vison) circles around $y$ direction will acquire an Aharonov--Bohm phase $\exp (i\theta c)$, where the charge of the quasiparticle is $c=1/2$. Thus,
\begin{equation}
\mathcal T_s^y \mathcal F_s^x(\theta)|\psi_0\rangle=\exp (i\theta c)  \mathcal F_s^x(\theta)|\psi_0\rangle.
\end{equation}
If one inserts $2\pi$ flux, the system will be brought back into the original system, however, due to the fractional charge, the Aharonov--Bohm phase from the flux will be $-1$. Therefore, $\mathcal F_s^x(2\pi)|\psi_0\rangle$ is just the state $|\psi_s\rangle$,  $\mathcal T_s^y \mathcal F_s^x(2\pi)|\psi_0\rangle=-\mathcal F_s^x(2\pi)|\psi_0\rangle$. In general, we have
\begin{align}
|\psi_s\rangle&=\mathcal F_s^x(2\pi)|\psi_0\rangle, \quad |\psi_v\rangle=\mathcal F_v^x(2\pi)|\psi_0\rangle, \nonumber \\  |\psi_{sv}\rangle&=\mathcal F_s^x(2\pi)\mathcal F_v^x(2\pi)|\psi_0\rangle. \label{eq:top_flux}
\end{align}
The topological degenerate ground states obtained by inserting flux only rely on the fact of fractionalized particles, there is no difference between Z$_2$ spin liquid and double-semion spin liquid.

On the other hand, we can deduce the topological degeneracy from the braiding statistics of quasiparticles. The anyonic braiding statistics of Z$_2$ spin liquid can be summarized as the following. Firstly, either spinon or vison has trivial braiding statistics to itself. Secondly, spinon (vison) obeys semionic  statistics relative to vison (spinon), which means if a spinon (vison) encircles around a vison (spinon), there will be a phase factor $-1$. The braiding statistics between spinon and vison in Z$_2$ QSL can be written as:
\begin{align}
\mathcal{T}_{s}^{x} \mathcal{T}_{v}^{y}&=-\mathcal{T}_{v}^{y}\mathcal{T}_{s}^{x}, \quad \mathcal{T}_{v}^{x} \mathcal{T}_{s}^{y}=-\mathcal{T}_{s}^{y}\mathcal{T}_{v}^{x}, \nonumber \\ \quad \mathcal T^x_s \mathcal T^y_s&=\mathcal T^y_s \mathcal T^x_s , \quad \mathcal T^x_v \mathcal T^y_v=\mathcal T^y_v \mathcal T^x_v.
\end{align}
Then,  $\mathcal{T}_{s}^y(\mathcal{T}_{s}^x|\psi_0\rangle)=\mathcal{T}_{s}^x\mathcal{T}_{s}^y|\psi_0\rangle=( \mathcal{T}_{s}^x|\psi_0\rangle)$ and $\mathcal{T}_{v}^y(\mathcal{T}_{s}^x|\psi_0\rangle)=-\mathcal{T}_{s}^x\mathcal{T}_{v}^y|\psi_0\rangle=-( \mathcal{T}_{s}^x|\psi_0\rangle)$. Therefore, $|\psi_v\rangle=\mathcal{T}_{s}^x|\psi_0\rangle$ is the ground state in the sector-$(1,-1)$. This sector can be simply understood as a state with a spinon line in the $x$ direction, as in Fig. \ref{fig:top_deg}(b). Similarly, we can have two other ground states. In general, in a Z$_2$ QSL we have:
\begin{equation}
|\psi_v\rangle=\mathcal T_s^x |\psi_0\rangle, \,\, |\psi_{s}\rangle=\mathcal T_v^x|\psi_0\rangle,  \,\, |\psi_{sv}\rangle=\mathcal T_s^x \mathcal T_v^x|\psi_0\rangle, 
\end{equation}
and, 
\begin{equation}
 \mathcal T_s^x=\mathcal F_v^x(2\pi), \quad \mathcal T_v^x=\mathcal F_s^x(2\pi). \label{eq:z2_flux_anyon}
\end{equation}

On the contrary, the double-semion QSL has quite different braiding statistics. Spinon (vison) obeys semionic braiding statistics to itself, but trivial braiding statistics to vison (spinon). The braiding statistics between spinon and vison in double-semion QSL can be written as:
\begin{align}
\mathcal{T}_{s}^{x} \mathcal{T}_{v}^{y}&=\mathcal{T}_{v}^{y}\mathcal{T}_{s}^{x}, \quad \mathcal{T}_{v}^{x} \mathcal{T}_{s}^{y}=\mathcal{T}_{s}^{y}\mathcal{T}_{v}^{x}\nonumber \\ \mathcal T^x_s \mathcal T^y_s&=-\mathcal T^y_s \mathcal T^x_s , \quad \mathcal T^x_v \mathcal T^y_v=-\mathcal T^y_v \mathcal T^x_v.
\end{align}
Therefore, in double-semion QSL we have:
\begin{equation}
|\psi_s\rangle=\mathcal T_s^x |\psi_0\rangle, \,\, |\psi_{v}\rangle=\mathcal T_v^x|\psi_0\rangle, \,\, |\psi_{sv}\rangle=\mathcal T_s^x \mathcal T_v^x|\psi_0\rangle,
\end{equation}
and,
\begin{equation}
 \mathcal T_s^x=\mathcal F_s^x(2\pi), \quad \mathcal T_v^x=\mathcal F_v^x(2\pi).  \label{eq:ds_flux_anyon}
\end{equation}

\section{Different topological sectors from infinite DMRG \label{sec:algorithm}} 
\begin{figure}[h]
\centering
\includegraphics[width=0.48\textwidth]{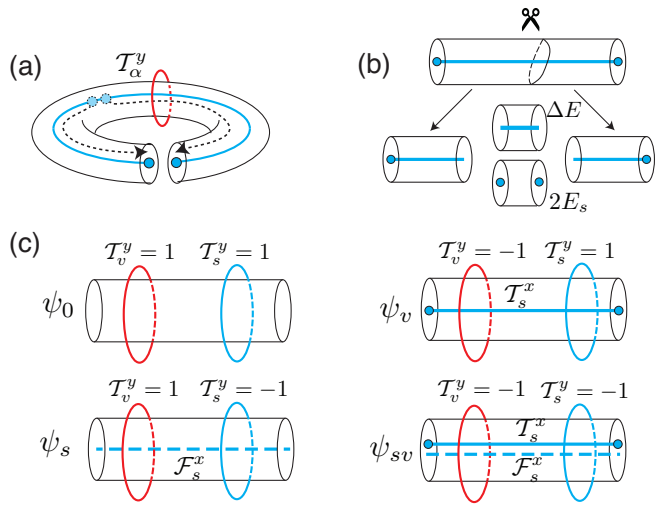} \caption{(color online) \label{fig:deg}(a) Cutting a torus into a cylinder. (b) Illustration of an infinite DMRG algorithm. (c)  $\psi_0$, $\psi_v$, $\psi_s$ and $\psi_{sv}$ of a Z$_2$ QSL from two operations $\mathcal T_s^x$ and $\mathcal F_s^x$.}
\end{figure}

The topological degeneracy is accurately defined on a torus, however one can also have topological degenerate sectors on an infinite cylinder. \cite{Cincio2013, Poilblanc2012, Schuch2013} This can be  understood by cutting the torus into a cylinder as shown in Fig. \ref{fig:deg}(a). The  subtle difference for the cylinder is that a pair of anyons will be standing at the two ends of the cylinder instead of annihilating each other as in the torus case. This  difference will not bring any distinct behaviors to the local wave function, as long as one measures the system far away from the ends. However, the existence of edge quasiparticles will bring large energy punishment to the system, as a result, different topological sectors only have the same energies in the bulk of the cylinder. 
This will not result in  any important  effect  in the infinite DMRG simulation, \cite{McCulloch2008} since one only optimizes the energy in the center of the infinite cylinder. 
A nice feature is that simulations on an infinite cylinder will collapse into one topological sector,
\cite{Cincio2013, Zaletel2013} which is  the key to the  controlled method we develop.

To obtain topological sectors systematically, we should find out how to realize those $\mathcal T^x_\alpha$ and $\mathcal F^{x}_\alpha$ operations. To drag out an anyon line, one can apply a string operator on the system,
however the string operator is usually difficult to identify.  
Here we use an equivalent way,  which does not drag out  the anyon line directly, but helps the system to develop an anyon line. 
This can be done by creating edge quasiparticles on the cylinder at the  beginning of  the infinite DMRG simulation as illustrated in Fig. \ref{fig:deg}(b).
In the simulation, one cuts the cylinder into two halves,  inserts one column of sites and optimizes the energy within the newly added sites.
This  procedure is repeated until the convergence is reached. It appears that  there are two options for the newly inserted sites, 
either with or without an anyon line attaching to them. If there is an anyon line, there will be an energy cost $\Delta E$, which comes from the energy splitting of different topological sectors due to finite size effect. On the contrary, if there is no anyon line, two quasiparticle domain walls will appear at the interface of newly added sites and the older system,
which  brings in an energy punishment $2E_q$ (the energy of two quasiparticles). Therefore,  if  $2E_q$ is much bigger than $\Delta E$, the simulation will collapse into the topological sector with an anyon line in it as desired. 

To apply the flux insertion operation $\mathcal F_\alpha^x$, we begin with DMRG simulation and obtain one topological sector first.
Now we can adiabatically turn on  the flux characterized by a boundary twist
parameter $\theta$ from $0$ to $2\pi$ gradually,
and the starting topological sector will evolve into another topological sector. 
The key point here is how to ensure the adiabaticity during the flux insertion. 
We discretize the flux $\theta\in (0,2\pi)$
and slowly increase $\theta$ to obtain new wave function by using the state from previous $\theta$
 as the initial wave function. 
Under this procedure,  the system  should  evolve adiabatically into the new topological sector
to avoid having  two quasiparticle domain walls  at the interface between the inserted sites and the 
older system.

In a QSL, there  exists at least one type of quasiparticles, the spinon, which can be considered as an unpaired spin in the RVB representation. Thus, to get a topological sector with a spinon line threaded in, we can simply put unpaired impurity spins (pin or remove one site) on the left and right boundaries similar to the pinning in the earlier work. \cite{Yan2011} Meanwhile, the spinon flux insertion can be realized by the twist boundary condition.\cite{Niu1985,Misguich2010}
In the hard-core boson representation of the spin system,
the spinon is the fractionalized hard-core boson with charge number $1/2$.
Then flux $\mathcal F_s$ is just the ``magnetic'' flux seen by  hard-core bosons, which can be realized by the twist  boundary condition along the $y$ direction:
\begin{equation}
S_1^+S_N^-+S_1^-S_N^+ \rightarrow e^{i\theta} S_1^+S_N^-+e^{-i\theta}S_1^-S_N^+.  \label{eq:twist}
\end{equation}
A similar twist technique has been  used in the 2D extension of the LSM theorem, \cite{Hastings2004}  however  there is a difference which was also mentioned in Ref. \onlinecite{Oshikawa2006}. The flux insertion in 2D LSM theorem relies on the momentum 
counting,
while our method applies to a topologically fractionalized phase
relying on the adiabatic evolution of the topological sector.

In a QSL, we can apply these two operations $\mathcal T_s^x$ and $\mathcal F_s^x$ to get different topological sectors.
Since only spinon is sensitive to the spinon flux $\mathcal F_s^x$, the topological sector obtained by spinon flux insertion should always have the Wilson loop operator $\mathcal T_s^y=-1$, as shown in Eq. (\ref{eq:top_flux}). On the contrary, which  topological sector we  obtain through developing spinon line will depend on the details of the quasiparticle braiding statistics in the QSL. 
In a Z$_2$ QSL, the spinons (visons) obey the mutual semionic braiding statistics relative to the
visons (spinons), but trivial statistics to themselves. Then dragging out a spinon line equals to inserting a vison flux (Eq. \ref{eq:z2_flux_anyon}), since only vison is sensitive to the spinon line. Therefore, the ground states in four sectors of a Z$_2$ QSL can be obtained by those two operations $\mathcal T_s^x$ and $\mathcal F_s^x$ as shown in Fig. \ref{fig:deg}(c). In contrast, for the double-semion QSL, \cite{Levin2003} the spinons (visons) obey semionic braiding statistics to themselves but trivial statistics relative to visons (spinons). Then, only spinon is sensitive to spinon line, so that  dragging out a spinon line is equivalent to inserting a spinon flux (Eq. \ref{eq:ds_flux_anyon}). Therefore, these two operations will lead to  exactly the same state and we need other operations to get new topological sectors.

\section{The winding number, spinon line and entanglement spectrum \label{sec:winding_number}}

In  literatures, the winding number is often used to label the topological sectors in the RVB type
QSL. The winding number $W_{x(y)}$ is defined by the parity of the number of the singlets cut by a loop along the $x$ ($y$) direction. In fact, a topological sector with a spinon line threaded in $x$ direction just has winding number $W_y=-1$. \cite{Misguich2010} To understand this, we consider a cylinder with even number of sites one column, and we do a cut along $y$ (vertical) direction. For $2N$ spins on the half cylinder, $2N_p$ of which are forming singlet pairs within the half cylinder, $N_c$ of which are forming singlet with spins from the other half cylinder. If there is no unpaired spinon in the cylinder, we have $2N=2N_p+N_c$, then $N_c$ is an even number, $W_y=1$. If there is a spinon line in the cylinder, leaving unpaired spinon on the left and right edge of the cylinder, then $2N=2N_p+N_c+1$, which means $W_y=-1$. 

Although we have an exact correspondence between spinon line and winding number, the form of the Wilson loop operator corresponding to the winding number actually depends on the details of quasiparticles braiding statistics. For example, in a Z$_2$ QSL, the winding number $W_y$ is the eigenvalue of Wilson loop operator $\mathcal T_v^y$, while in the double-semion QSL, $W_y=\mathcal T_s^y$. 
In a Z$_2$ QSL, the topological sectors labeled in winding number $(W_x=\pm 1, W_y=\pm 1)$ can be represented by the superposition of  the topological sectors we used in the paper: 

\begin{align}
|\pm , +\rangle&= |\mathcal T_s^y=1, \mathcal T_v^y=1\rangle \pm |\mathcal T_s^y=-1, \mathcal T_v^y=1\rangle  \\  |\pm ,-\rangle&= |\mathcal T_s^y=1, \mathcal T_v^y=-1\rangle \pm |\mathcal T_s^y=-1, \mathcal T_v^y=-1\rangle. \nonumber
\end{align}
However,  for some QSLs those two winding numbers are not enough to label all the topological sectors. 
On the other hand,  for some QSLs like the double-semion state, where spinons have the
nontrivial braiding statistics relative to themselves, these two winding numbers can not be used simultaneously to label the topological sectors since the corresponding operators do not commute  with each other.

There is a simple way to judge whether a state has a spinon line in it by  observing the symmetry properties of the entanglement spectrum (ES) corresponding to a vertical-loop cut. For the sector with no spinon line, the ES  is symmetric about total $S^z=0$; while for the sector with a spinon line, the ES is symmetric about $S^z=1/2$. The reason for this different symmetry behavior simply comes from the even or odd number of singlets cut by the loop. In the RVB states, only the singlet pair being cut contributes  a $\pm 1/2$
to the total $S_z$ with equal probability.  As a result, ES of the state with the different winding number will have different symmetry (Fig. \ref{fig:entanglement}).

\section{Easy axis kagome model \label{sec:easy_axis}}
\begin{figure}[ht]
\centering
\includegraphics[width=0.48\textwidth]{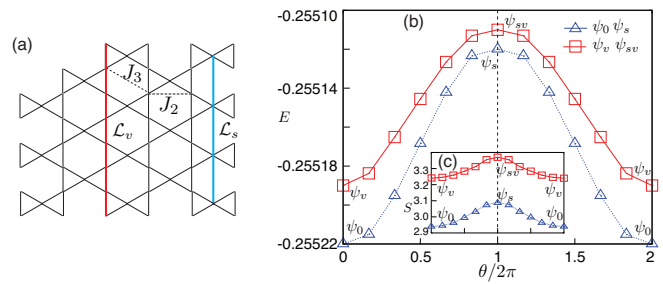} \caption{(color online)\label{fig:easy-axis} (a) Easy axis kagome with second and third nearest interaction. The loop operators $\mathcal L_v$ and $\mathcal L_s$ are defined as $\mathcal L_v=\prod 2S^i_z$ and $\mathcal L_s=\prod 2S^i_x$.  Energy (b) and entropy (c) evolution of four topological sectors under flux insertion. The parameter we take here is $J^{xy}=-0.1$.}
\end{figure}

We apply our method to the easy axis kagome system, which has a  Hamiltonian in the following form:
\begin{equation}
H=\sum J _{ij}^z S_i^z S_j^z+ \sum \frac{J_{ij}^{xy}}{2} (S_i^+ S_j^-+S_i^-S_j^+),
\end{equation}
where the spin $z$ components  have the first, second and third nearest neighbor coupling terms with the same magnitude (Fig. \ref{fig:easy-axis}(a)), while the  spin
 $x$ and $y$ components  have only the first nearest neighbor couplings. This model is shown to host a Z$_2$ QSL both theoretically \cite{Balents2002} and numerically \cite{Sheng2005a,Isakov2006,Isakov2011} for $J^z_1=J^z_2=J^z_3=J^z=1$ and FM $J_1^{xy} \geq -0.14$.
\cite{Isakov2011} The system  we study has $4$ unit cells (8 lattice sites)
one column.

We obtain the topological sector $\psi_0$ naturally in the conventional DMRG simulation, and find the  $\psi_v$ by creating the edge spinon as discussed before. 
Then we apply the adiabatic flux insertion to get the other two sectors. 
 The evolution of the energy and entropy
  with flux insertion  is  demonstrated  in Fig. \ref{fig:easy-axis}.  
Apparently,  by threading the $2\pi$ flux, the two starting states $\psi_0$ and $\psi_v$ both evolve into the new ground states ($\psi_s$ and $\psi_{sv}$)
in other topological sectors, and the states evolve  back to the original states after inserting the $4\pi$ flux.

\begin{figure}[ht]
\centering
\includegraphics[width=0.48\textwidth]{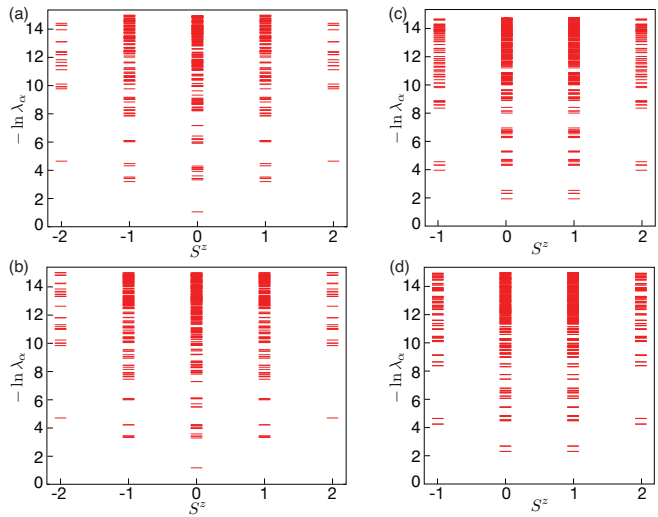} \caption{\label{fig:entanglement} Entanglement spectrum of four Z$_2$ sectors. (a) $\psi_0$ (b) $\psi_s$ (c) $\psi_v$ (d) $\psi_{sv}$. Since we are using a $U(1)$ symmetry matrix product state representation, each entanglement spectrum has an $S^z$ charge. Here $J^{xy}=-0.1$.}
\end{figure}

Entanglement spectrum (ES) of four states are shown in Fig. \ref{fig:entanglement}: the state ($\psi_0$, $\psi_s$) without spinon line has ES symmetric about $S^z=0$; the state  ($\psi_v$, $\psi_{sv}$) with spinon line has ES symmetric about $S^z=1/2$. 

To verify that  these four states are distinct states from different Z$_2$ topological sectors, we define and measure the loop operators $\mathcal L_v$ and $\mathcal L_s$ as shown in Fig. \ref{fig:easy-axis}(a). 
At the RK exactly solvable point, \cite{Balents2002} $\mathcal L_v$ and $\mathcal L_s$  will be identical to the Wilson loop operators $\mathcal T_v$  and $\mathcal T_s$, respectively. 
The obtained expectation values of the  $\mathcal L_v$ and $\mathcal L_s$ for different states are shown in Table \ref{table:wilson}. Clearly, in our parameter region, $\mathcal L_v\approx \pm 1$ is still a very good operator to approximate  $\mathcal T_v$.  On the other hand, the expectation values of $\mathcal L_s$ for the different states are much smaller than $1$, but they still are reasonably large values compared to the zero. 
More importantly, the two loop operators take distinct signs in four sectors, which fit well with the Z$_2$ QSL theory illustrated in Fig. \ref{fig:deg}(c).

\begin{table}[h]
\centering
\begin{tabular}{c||ccccccccc}
\hline
$J^{xy}$&state & $\mathcal L_v$ & $\mathcal L_s$ & $E$ & $S$ & $\xi_{\textrm{TM}}$ & $\langle\psi_0|$& $\langle\psi_s|$& $\langle\psi_v|$  \\ \hline  \multirow{4}{*}{$-0.1$} &$|\psi_0\rangle$& 0.90 & 0.21 & -0.25522 & 2.94 & 0.63 & & &
 \\  
 & $|\psi_s\rangle$ & 0.91 & -0.18 & -0.25512 & 3.09 & 0.79 &0.49 & & \\ 
 &$|\psi_v\rangle$ & -0.90 & 0.13 & -0.25519 & 3.24 & 0.74 & 0.39&0.19 &  \\  
 & $ |\psi_{sv}\rangle$ & -0.90 & -0.12 & -0.25511 & 3.37 & 0.9 &0.18 &0.29 & 0.54\\ \hline \hline
 \multirow{4}{*}{$-0.05$} &$|\psi_0\rangle$& 0.98 & 0.20 & -0.251199 & 2.94 & 0.65 & & &
 \\ 
 & $|\psi_s\rangle$ & 0.98 & -0.20 & -0.251197 & 2.97 & 0.66 &0.43 & & \\ 
 &$|\psi_v\rangle$ & -0.98 & 0.12 & -0.251194 & 3.30 & 0.82 & 0.14&0.06 &  \\  
 & $ |\psi_{sv}\rangle$ & -0.98 & -0.11 & -0.251192 & 3.32 & 0.84 &0.06 &0.12 & 0.52\\ \hline
\end{tabular}
\caption{ The loop operators, energy, entropy, correlation length $\xi_{\textrm{TM}}$ and one column overlap $\chi_{ij}$ of four topological sectors. The unit of correlation length is unit cell.}
\label{table:wilson}
\end{table}

\begin{figure}[ht]
\centering
\includegraphics[width=0.48\textwidth]{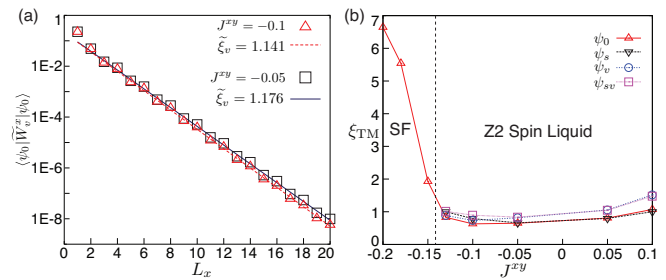}\caption{(color online) \label{fig:correlation} (a) The vison-vison correlation length $\widetilde \xi_v$ from $\langle \psi_0|\widetilde W_v^x |\psi_0\rangle$. (b) $J^{xy}$ dependent behavior of four topological sectors' correlation lengths. For $J^{xy}\lesssim -0.14$, the system is in the superfluid (SF) phase, for $J^{xy}\gtrsim -0.14$, the system is in the Z$_2$ spin liquid phase, where one has four topological degenerate ground states.}
\end{figure}

The correlation length $\xi_{\textrm{TM}}$ measured from the transfer matrix (Append. \ref{sec:num}) for each of these states
is much smaller than the cylinder width indicating a small finite size effect.
We also show the one column overlap $\chi_{ij}$ between states in different sectors (Append. \ref{sec:num})  in Table \ref{table:wilson}. The overlap we calculate here is conceptually different from the overlap in  a finite system. If we consider the overlap between different sectors on a cylinder with length $L$, then we will obtain $(\chi_{ij})^L$, which vanishes exponentially fast. On the other hand, since the 
different topological sectors only differ from each other due to the different types of the quasiparticle lines threaded in them, the overlap between these sectors on a cylinder is just  the spinon-spinon or vison-vison correlation:
\begin{align}
\langle \psi_0|\mathcal S(0) \mathcal S(L)|\psi_0\rangle &=\langle \psi_0 | \psi_v\rangle=(\chi_{0v})^L, \,\xi_s=-1/\ln \chi_{0v}, \nonumber \\
\langle \psi_0|\mathcal V(0) \mathcal V(L)|\psi_0\rangle& =\langle \psi_0 | \psi_s\rangle=(\chi_{0s})^L, \, \xi_v=-1/\ln \chi_{0s}.
\end{align}
Here $\mathcal S(0)$ and $\mathcal S(L)$ ($\mathcal V (0)$ and $\mathcal V(L)$) represent the spinons (visons) located at the left and right edge of the cylinder, respectively. The $\xi_s$ and  $\xi_v$ are the spinon-spinon and the vison-vison correlation lengths, respectively. We can also extract the vison-vison correlation length approximately from $\langle \psi_0|\widetilde {\mathcal L_v^x}|\psi_0\rangle$, where $\widetilde  {\mathcal L_v^x}=\prod 2S_i^z$ is a string operator defined in the $x$ direction, similar to  the loop operator $\mathcal L_v$ shown in Fig. \ref{fig:easy-axis}(a). Extrapolating $\langle \psi_0|\widetilde  {\mathcal L_v^x}|\psi_0\rangle$ (Fig. \ref{fig:correlation}(a)), we get the vison-vison correlation length, $\widetilde \xi_v=1.141$ for $J^{xy}=-0.1$, and $\widetilde \xi_v=1.176$ for $J^{xy}=-0.05$. On the other hand, the vison-vison correlation length from the overlap $\chi_{0s}$ is $\xi_v=1.402$ for $J^{xy}=-0.1$ and $\xi_v=1.185$ for $J^{xy}=-0.05$. At $J^{xy}=-0.1$, two correlation lengths have a bigger difference because
  the string operator $\widetilde  {\mathcal L_v^x}$  starts to have a considerable deviation from the real vison-vison correlation operator for more negative $J^{xy}$. The spinon (vison) correlation can also be extracted from the energy splitting $\Delta E$ of different topological degenerate states by $\Delta E_{s(v)}\sim \exp (-L_y/\xi_{s(v)})$. \cite{Schuch2013}
  
To see the system's behavior under different $J_{xy}$, we plot the correlation length in Fig. \ref{fig:correlation}(b). For $J^{xy}\lesssim-0.14$, we only get one ground state and it has a very large correlation length indicating a gapless superfluid state; $J^{xy}\gtrsim -0.14$, the four topological degenerate ground states with degenerating bulk energies always exist.

\section{Kagome Heisenberg model \label{sec:KHM}}
In this section, we will discuss  the kagome Heisenberg model with the following Hamiltonian:
\begin{equation}
H=\sum_{\langle i, j\rangle} \bm S_{i}\cdot \bm S_j+J_2\sum_{\langle \langle i, j\rangle\rangle} \bm S_{i}\cdot \bm S_j.
\end{equation}
Here the spins are coupled by the isotropic nearest neighbor $J_1$ and the next nearest neighbor $J_2$ interactions. In the following, the simulation is mainly done on systems with a width of $4$ unit cells ($8$ lattice sites) and  for convenience we stick to the notation of a Z$_2$ QSL  used before, which we find is more consistent with what we have observed for this system.  

\begin{figure}[ht]
\centering
\includegraphics[width=0.48\textwidth]{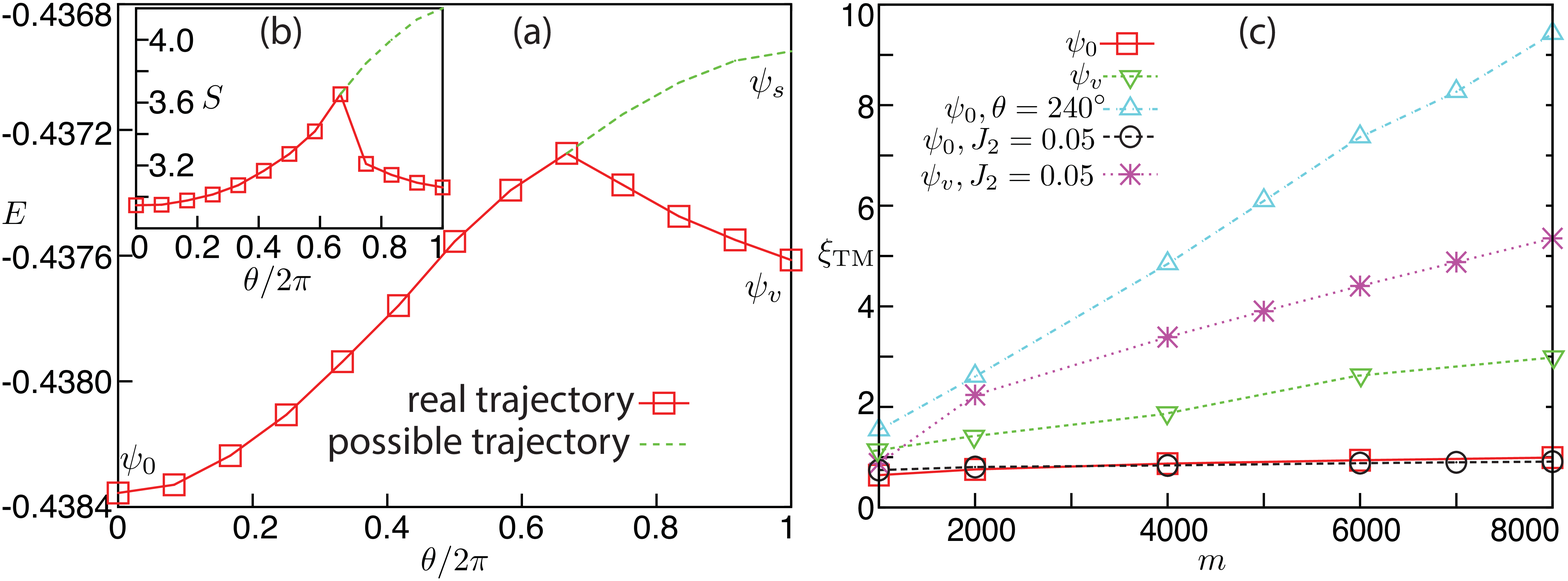} \caption{(color online)\label{fig:kagome}  Energy (a) and entropy (b) evolution of $\psi_0$ at $J_2=0$. (c) Correlation length $\xi_{\textrm{TM}}$ of different topological sectors versus the states kept. Here the unit of the correlation length is the unit cell.}
\end{figure}

Without and with creating edge spinons we can get the two states $\psi_0$ and  $\psi_v$, and the energy splitting  between
these two states for $J_2=0$ is $0.00070(4)$ per site, which is consistent with the results of the Ref. \onlinecite{Yan2011}.
To get the other two topological sectors, we apply the  twist boundary phase (flux insertion) adiabatically, but we find that
 the adiabaticity of the evolution for the ground state  breaks down  around $\theta=240^\circ$ as shown in Fig. \ref{fig:kagome}(a) \cite{footnote2}. However one observation from our results is that the energy and the entropy continue to increase as the twist angle passes
 the time reversal invariant point $\theta=\pi$, which may indicate the existence of a different topological sector $\psi_s$. 
The breakdown  of the adiabaticity may result  from the instability of the $\psi_s$ as a higher energy state with larger splitting $\Delta E$. 

Although we can not fully access $\psi_s$, we think that the state $\psi_0(\theta=240^\circ)$ is close to $\psi_s$ at $\theta=360^\circ$, from which we can have some estimate of the properties of $\psi_s$. For $J_2=0$, the energy splitting between $\psi_s$ and $\psi_0$ should be larger than $0.00108$ and $\psi_s$ has an entropy larger than $3.6$. We also measure the correlation length from the transfer matrix,
as shown in Fig. \ref{fig:kagome} (b). The $\psi_0$ has a small correlation length, which is consistent with the results obtained in Ref. \onlinecite{Yan2011, Depenbrock2012}.  However, we find that the correlation lengths of other topological sectors, $\psi_v$ at $J_2=0$, $0.05$ and $\psi_0(\theta=240^\circ)$, are much larger. In these three states, the nearest bond spin-spin correlations are very homogeneous (Append. \ref{sec:bond}), so we think they are also the QSL states with no symmetry breaking.
Recently,  a gapless QSL has been explicitly constructed in Ref. \onlinecite{Wang2013}, 
 and one finds that in  a cylinder, some topological sector has a small correlation length while other topological sectors have a very large correlation length. This scenario is similar to  what we have obtained here. However, based on our numerical results, we can not draw a conclusion about
 the nature of QSL on the KHM due to the strong finite size effect associated
with longer correlation length, which  we leave for the future study.

\section{Summary}
A controlled method for DMRG approach to find  different topological sectors of the QSL on an infinite cylinder is proposed and applied to the  EAKM and  KHM. In EAKM,  the complete set of four Z$_2$ topological sectors are obtained. In KHM, we can get two topological sectors exactly, and estimate the properties of other topological sectors. We find larger correlation lengths for other topological sectors. Our numerical scheme based on  creating   boundary quasiparticles or threading  the  flux adiabatically, may also be applied to other numerical algorithms like the tensor network. Our method may help to understand the topological nature of different  QSL systems and bring new excitement to the study of QSLs.

\section{Acknowledgments}
This work was supported by the State Key Programs of China (Grant Nos. 2012CB921604 and
2009CB929204) and the National Natural Science Foundation of China (Grant Nos. 11074043 and 11274069)
(YCH and YC), and the US National Science Foundation  under grant DMR-0906816 (DNS).

\appendix
\section{Numerical Algorithm \label{sec:num}}
\begin{figure}[h]
\centering
\includegraphics[width=0.48\textwidth]{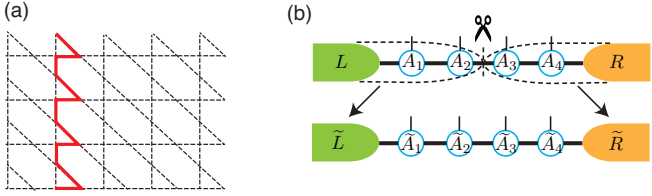} \caption{(color online)\label{fig:iDMRG}  (a) Covering a cylinder with one dimensional path. (b) Illustration of iDMRG algorithm with 4 sites in the column.}
\end{figure}

The numerical algorithm we use is the infinite density matrix renormalization group (iDMRG) invented by McCulloch. \cite{McCulloch2008} Similar to the finite DMRG, we use a one dimensional path to cover 
all the sites of the 2D cylinder. In the iDMRG calculation, we first get left and right Hamiltonian ($L, R$ in Fig. \ref{fig:iDMRG}(b)) from small size simulation (or choosing random initial state \cite{Cincio2013}). Then  we insert one column in the center and optimize the energy only within the inserted column by sweeping. After the optimization, we cut one column into two halves, absorb them into the left and right Hamiltonian respectively to get the new boundary Hamiltonians $\widetilde L, \widetilde R$, respectively. 
The inserting, optimizing and cutting procedure is repeated until the convergence is achieved. With the converged results for the column, one can represent the translational invariant wave function of the infinite cylinder or mimic  the wave function on a torus. \cite{Cincio2013}

\begin{figure}[h]
\centering
\includegraphics[width=0.48\textwidth]{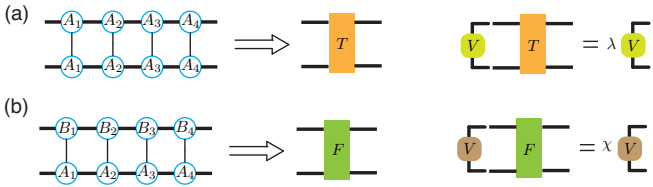} \caption{(color online)\label{fig:transfer}  (a) Calculating the correlation length $\xi_{\textrm{TM}}$. (b) Calculating the overlap between different states.}
\end{figure}

In the simulation of different topological sectors, there are two important quantities to calculate, the  correlation length $\xi_{\textrm{TM}}$ and overlap between different topological sectors. To calculate the correlation length $\xi_{\textrm{TM}}$, we should calculate the first and second largest eigenvalue $\lambda_{1,2}$ of the transfer matrix $T$ defined in Fig. \ref{fig:transfer}(a). For a normalized wave function, the largest eigenvalue $\lambda_1=1$. Then the correlation length $\xi_{\textrm{TM}}=-1/\ln \lambda_2$. This correlation length determines the largest correlation in the infinite cylinder \cite{McCulloch2008, Cincio2013}. Therefore, instead of calculating various correlation functions, one can simply calculate this single quantity $\xi_{\textrm{TM}}$ to know the length scale of the largest possible correlations. 

Similarly, the overlap of different states is the largest eigenvalue ($\chi$) of $F$ matrix defined in Fig. \ref{fig:transfer}(b). This overlap is slightly different from the overlap on a finite system. However, if one puts the wave function from iDMRG simulation on cylinder or torus with length $L$, then the overlap between two different states is $\langle \psi_1|\psi_2\rangle\approx\chi^L$.

\section{The bond spin correlation of the Kagaome Heisenberg Model\label{sec:bond}}

\begin{figure}[ht]
\centering
\includegraphics[width=0.48\textwidth]{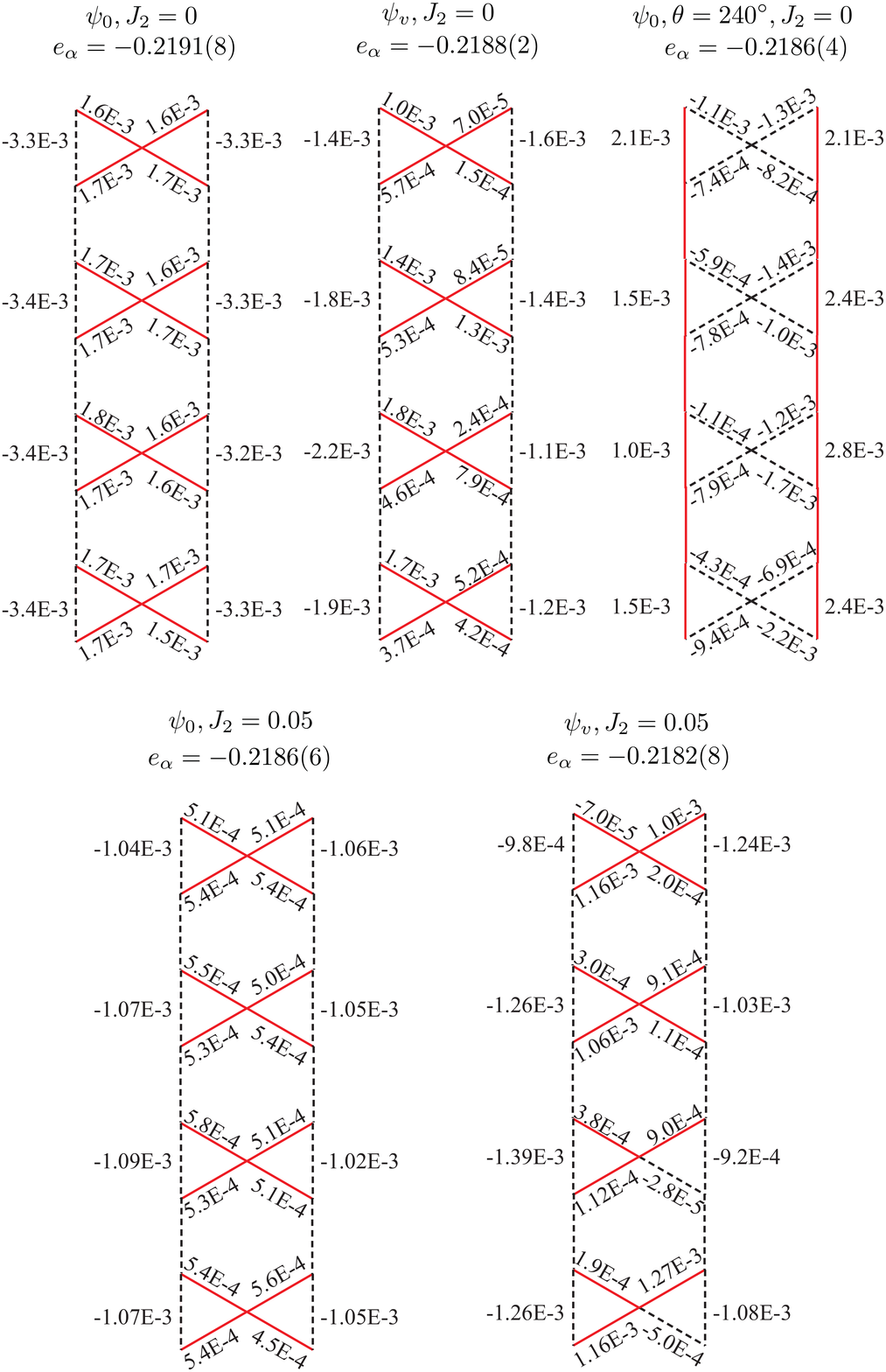} \caption{(color online)\label{fig:bond_energy}  The bond spin correlation $\langle \bm S_i \cdot \bm S_j\rangle-e_\alpha$ of different sectors, $e_\alpha$ is the average of bond spin correlations. All the results plotted are obtained by keeping 8000 states in DMRG simulations. The truncation error of $\psi_0$ is smaller than $10^{-6}$, while other states has a truncation error smaller
than  $5\times 10^{-6}$.}
\end{figure}

To know whether the obtained states
have lattice symmetry breaking, we need to check if the bond correlation is uniformly distributed in the whole system, which is shown  in Fig. \ref{fig:bond_energy}. 

For the 
state with nonzero twist, one should transform the twisted Hamiltonian into a translational invariant one by a gauge transformation,
\begin{equation}
U(\theta)=\prod_{x} \prod_{y=1}^N \exp(i\frac{y }{N} \theta S_{x,y}^z).
\end{equation}
Under the gauge transformation, the Hamiltonian becomes
\begin{align}
 J_z S_{x_1,y_1}^z S_{x_2,y_2}^z&+\frac{J}{2} (S_{x_1,y_1}^+S_{x_2,y_2}^-+S_{x_1,y_1}^-S_{x_2,y_2}^+) \nonumber \rightarrow \\  J_z S_{x_1,y_1}^z S_{x_2,y_2}^z&+\frac{J}{2} \left[e^{i(y_1-y_2)\theta/N} S_{x_1,y_1}^+S_{x_2,y_2}^- \right.  \nonumber \\  & \left.
 +e^{-i(y_1-y_2)\theta/N}S_{x_1,y_1}^-S_{x_2,y_2}^+ \right]. 
  \label{eq:hoppingphase}
\end{align}

The bond spin correlations in all the states are very homogeneous with  fluctuations around or smaller than $1\%$.


\begin{thebibliography}{99}
\bibitem{footnote2} We have tried many tricks (including using a very small step $2^\circ$) to ensure the adiabaticity, but all failed. The state flows into $\psi_v$ after the failure of adiabaticity is a random phenomenon. When the $J_2$ interaction is large $J_2\sim 0.1$, the state will flow back into $\psi_0$.
\end{thebibliography}
\end{document}